\documentstyle[12pt]{article}

\begin{document}
\begin{titlepage}

\title { QUANTUM WAVEGUIDE TRANSPORT IN SERIAL STUB AND LOOP
STRUCTURES }
\author{P. Singha Deo$^1$ and A. M. Jayannavar$^2$
\\
Institute of Physics, Bhubaneswar-751005, INDIA.}

\footnotetext[1]{e-mail:prosen@iopb.ernet.in}
\footnotetext[2]{e-mail:jayan@iopb.ernet.in}

\maketitle

\thispagestyle{empty}

\begin{abstract}

We have studied the quantum transmission properties of serial
stub and loop structures. Throughout we have considered free
electron networks and the scattering arises solely due to the
geometric nature of the problem. The band formation in these
geometric structures is analyzed and compared with the
conventional periodic potential scatterers. Some essential
differences are pointed out. We show that a single defect in an
otherwise periodic structure modifies band properties non
trivially. By a proper choice of a single defect one can produce
positive energy
bound states in continuum in the sense of von Neumann and Wigner.
We also discuss some magnetic properties of loop structures in
the presence of Aharonov-Bohm flux.

\end{abstract}

\end{titlepage}

\eject

\newpage
\hspace {0.5in}  In the past decade remarkable advances have taken
place in
micro fabrication, and now it is possible to confine electrons
in a conductor with a lateral extent of 100 nm or less,
resulting in narrow quantum wires, constrictions and quantum
dots[1-4]. This expansion
is due to combination of new developments in
lithography, patterning and layer growth techniques. Specifically it
is now possible to engineer device potentials which vary over
the length
scale such that the electron motion is ballistic or quasiballistic
at low temperatures.
The small size of these structures largely eliminate the
defect scattering and one can get extremely high mobility
conductive channels, thus motivating interest in the ballistic
regime. In these mesoscopic systems electron transport
is governed by quantum mechanics rather than classical
mechanics. At very low
temperatures, the scattering of phonons (dephasing scattering) is
significantly suppressed and the
phase coherence length of the electrons $L_{\phi}$ (the length
over which the electron can be considered to be in a pure
state),
becomes large compared to the system dimensions.
Mesoscopic system can
thus be modelled as a phase coherent elastic scatterer. The
idealized sample becomes an electron waveguide where the
transport properties are solely determined by the impurity
configuration, the geometry of the conductor and by the
principles of quantum mechanics.
As the phase coherence is maintained over the
entire sample, several
intrinsic quantum mechanical phenomenon have been observed[1-4].
This
has opened a completely new branch of device physics and
mesoscopic physics. Convincing
demonstration of new quantum transport regime have come from
experiments on thin metal or semiconductor films and multiply
connected structures. Some of the observed quantum phenomenon include
Aharonov-Bohm oscillations in the magneto resistance in doubly
connected ring structures[5], universal conductance fluctuations,
reproducible sample specific non-self-averaging fluctuations in
conductance as the magnetic field or the chemical potential is
varied, quantized conductance in the point contact, quenching of
Hall resistances in narrow cross
etc[2,3]. Given the coherence of the electrons throughout
the sample, several exciting ideas
for active quantum devices(transistors, switches etc) have been
proposed based on interferometric principles[6]. These are quantum
analog of well known optical or microwave devices. In these
structures, the electron transport is identical to the microwave
propagation through a wave guide.

\par In recent years, several studies have been reported on the
transmission across a T shaped device
consisting of a main wire of constant width attached to a
stub perpendicular to the wire (fig. 1)[7-11].
In this structure the
transmission
oscillates between zero(antiresonance) and one(resonance) with kL.
Here k is
the incident wave vector and L is the length of
the stub. Especially for the case of one channel T shaped
structures one obtains results similar to  the wave guides
resonantly coupled to a cavity[9,10]. Here the cavity has a set
of resonant states given by kL=n$\pi$. Along the main wire
across the stub,
transmission zeros (total
absence of forward scattering or total backscattering) and
resonances occur at $kL=n\pi$ and $kL=(n+1/2)\pi$ respectively.
The existence of transmission zeros are very specific to the
geometric nature of the scattering and do not occur in conventional
one dimensional potential scattering problems. The resonantly coupled
quantum waveguides also exhibits the zero-pole pairs in the
structure of transmission
amplitude in the complex energy plane[9,10].
The proximity of the zero and
the poles lead to sharp variations of transmission coefficients as a
function of energy and in certain circumstances lead to asymmetrical
Fano resonances[12]. If now
a multiple stub configuration is taken, the sharp drops to
zero in transmission become extended to
forbidden bands and the round tops get squared, along with
resonances to form allowed
bands. In ref [7,8] the potential usefulness of such systems
as transistor device (quantum modulated transistor) has been
discussed.
In this the drain current can be modified by remote
control where the
non-locality of electron waves is exploited, i.e.,
the
transistor action can be achieved by varying the effective length of
the vertical terminated lead. Relatively small changes in the
stub length can introduce strong variations in the electron
transmission across the structure.
However, quantum channels have a very high resistance($h/e^{2}$ =
25 k$\Omega$) and the current through a quantum channel is very
small. This practical problem can be resolved by stacking
superlattice of many such channels[8]. Other potential
switching devices are based on controlling the relative phase between
different interfering paths ( say in semiconducting loop structures)
by applying electrostatic or magnetic fields[6,13]. These
ballistic
devices promise to be much faster and will consume less power
than the conventional devices. The conventional
transistors operate in a classical diffusive regime and are not
sensitive to variations in material parameters such as dimensions or
presence of small impurities. These devices operate by controlling
the carrier density of quasiparticles. However, proposed quantum
devices are not very robust in the sense that the operational
characteristics depend very sensitively on material parameters. For
example, incorporation of a single impurity or slight structural
variations in the mesoscopic device
can change non-trivially the influence of partial waves propagating
through the sample and hence electron transmission across the sample.
In other words the electron transmission across these devices
is very specific to the
arrangement of the elastic scatterers
and on the Fermi energy. These devices can be exploited, if
we can achieve the technology that can reduce or control the phase
fluctuations to a small fraction of 2$\pi$. On the positive side it
should be noted that quantum devices can exhibit multifunctional
property (e.g, single stage frequency multiplier), wherein it can
perform the functions of an entire circuit within a single
element[14].

\par In this paper we discuss in detail many features of
band structure in multiple stub and ring (in the presence of magnetic
field) configurations. Throughout our analysis we have restricted
ourselves to one channel case, in that the main wire as well as the
stub is taken to be one dimensional. This single channel case
provides a good approximation to a
real wire with finite width at low
temperatures at which only the lower subband is filled.
Moreover, energy level spacings produced by transverse
confinement must be larger than the energy range of the
longitudinal transport and thermal broadening $k_{B}T$. In this
regime quantum wire behaves as a single moded electron waveguide. It
is also well known that to have a optimum performance, the quantum
device has to be operated in a fundamental mode, i.e., the Fermi
energy should be between the ground and the first excited transverse
mode. In this regime one can achieve a sharp and controllable
modulation of the electron transmission probability. With higher
energies (larger than the first excited transverse modes), the
propagation of various modes become possible and due to mode mixing
the total transmission becomes very sensitive to the  defect
structures and oscillations in the total transmission
coefficient are averaged out. Thus the energy dependence of the
total transmission becomes aperiodic.
In the case of single channel one can
apply quantum waveguide theory on networks[15-17]
and one can easily obtain
the transmission and reflection probabilities analytically.
In our analysis we have considered free electron networks, i.e.,
when
quantum potential V throughout the network is assumed to be
identically zero. The scattering arises solely due to junctions (or
geometric scattering) in free electron networks. The scattering
properties of the network
can be calculated by using the Griffith's
boundary conditions at the junction points[18].
These boundary conditions
are due to the single valuedness of the wave functions and the
conservation of the current(Kirchoff's law) at the junction. The
electron wave function has to be single valued throughout the
network. If
i segments intersect to form a junction and $\phi_{1}$,
$\phi_{2}$...$\phi_{i}$ are the wavefunctions on the segments,
the boundary condition at the junction point are
$\phi_{1}=\phi_{2}=....=\phi_{i}$ (continuity) and $\Sigma_{i}
{\partial \phi_{i}\over \partial x}$=0 (current conservation).
Here all the derivatives are either outward or inward from the
junction.
We have considered both serial stub structure (figs.
(1) and (2)) and
loop structures (figs. (3) and (4)).
For a single stub of length L as shown in fig. (1)
and for a loop structure as shown in fig. (3)
with equal arms of length L each and in the presence of magnetic
flux $\phi$,
Xia has obtained an analytical expression for the
transmission coefficient $\tau_{s}$ and $\tau_{l}$, respectively
and are given by[15]
\begin{equation}
\tau_{s}={4sin^2(kL)\over 4sin^{2}(kL)+cos^2(kL)}
\end{equation}

\begin{equation}
\tau_{l}={16(1-cos(2kL))(1+cos(\alpha))\over(1+4cos(\alpha)
-5cos(2kL))^2+(4sin(2kL))^2}
\end{equation}
where $\alpha=2\pi\phi/\phi_{0}$.
The eqn. (1) is for a single stub structure, where k is the
incident wave vector of an electron and L is the length of the
stub. In eqn. (2), L is the length of the single arm of the
loop. The two arm lengths are assumed to be equal, $\phi$ is the
magnetic flux and $\phi_{0}$=hc/e is the flux quantum. As
expected the transmission coefficient is flux periodic for all
energies with a period $\phi_{0}$.
For the expressions of complex reflection and
transmission amplitudes we refer to [15]. With the help of these
amplitudes one can easily compute the transmission coefficient of
regularly placed stub structure (fig. (2)) or ring structure
(fig. (4))
with the help of transfer matrix method. We have obtained an
analytical
result for these cases. The stubs (or rings)
are assumed identical and are placed at regular intervals of
length l. Coherent interference effects, due to elastic
scattering by serial structures can produce broad regions
where transmission is vanishingly small and these regions are
called conduction gaps (or forbidden gaps) and broad regions
with finite transmission along with resonances
and these regions are called conduction
bands.
We compare the evolution of band features
to that of band structure
in a one dimensional periodic potential scatterers, as
a function of the number of scatterers, and
point out some notable
differences. We next consider the effect of a
single
defect in an otherwise periodic structure. In particular we show
that a
single defect stub can change non-trivially the band structure.
A single defect stub in such a multiple stub
configuration can produce one or more zero transmission in the
conduction bands.
The length of the defect stub determines precisely where the
transmission
zero will be formed and adjusting the length we can adjust the
zeros (band tailoring). These dramatic changes in the band
structure are associated with the existence of transmission zeros
in geometric structures, which are absent in the conventional
potential scattering problems.
With an appropriate choice of the length of a single defect
stub, one can produce bound states in
continuum in the sense of von Neumann and Wigner[19].
Finally we have studied
the band formation due to a periodic structure of
mesoscopic loops with a Aharonov-Bohm flux passing through the
center of each loop and the effect of magnetic field on
the band evolution is studied. We briefly remark on the
paramagnetic and diamagnetic (orbital currents) properties of
the bands.
\par {\bf TRANSMISSION IN PERIODIC STRUCTURES}
\par Consider a serial structure shown in fig. (2). We can
disassemble the system into basic pieces connected in series.
The basic piece is the structure as shown in fig. (1).
We consider the system to be one dimensional
only to simplify the problem without any loss of
generality. If we consider the system to be of higher dimensions
(i.e., wires with finite transverse widths) then for small
values of incident electron energy (such that
there is only one propagating channel), the conductance
versus energy plot is similar to that of a 1-d system only the
origin being shifted due to the zero point energy of the lowest
transverse mode. To get the electron wave function in the
wire we have to solve the Schr$\ddot o$dinger eqn.
\begin{equation}
\nabla^2\phi+(E-V(x))\phi=0,
\end{equation}
we have set throughout the
units of $\hbar$ and 2m to be unity. In our case V=0
and the solutions are the
plane wave solutions i.e., $e^{ikx}$, where $k=\sqrt{E}$.
Then we adopt a technique that is specially
suited to such a serial structure. By the mode matching
mechanism we derive a transfer matrix that relates the
coefficients of the wave functions
at one end to that at the other end of the basic
piece. The total transfer matrix is just the product of the
transfer matrices of all the basic pieces in order[20].
\par The transfer matrix of the geometry shown in fig. (1) is
\begin{equation}
T^{(1)}=\pmatrix{{1/t^*} & {r/t} \cr
{r^*/t^*} & {1/t}}
\end{equation}
\noindent where $t$ is the complex transmission amplitude,
$r$ is the reflection amplitude and * denotes the complex
conjugation. These coefficients
can be determined by simply matching the boundary
conditions. Using the Griffith boundary conditions
at the junction and setting the wavefunction at the
free end of the stub to be zero (hard wall
boundary condition),
we get the transfer matrix due to one such
stub as[15]

\begin{equation}
T^{(1)}=\pmatrix{{(2-i\cot(kL))/2} & {-(i\cot(kL))/2} \cr
{(i\cot(kL))/2} & {(2+i\cot(kL))/2}}\,\,\,  ,
\end{equation}

\noindent where $L$ is the length of the stub. For a periodic stub
structure we have taken length of all the stubs to be equal and
they are placed at a length l apart as shown in fig. (2).

The total transfer matrix of N stub system is given by[20]

\begin{equation}
T=T^{(N)}MT^{(N-1)}MT^{(N-2)}M..............T^{(1)}\,\,\,  ,
\end{equation}
where
\begin{equation}
M=\pmatrix{{e^{ikl}} & {0} \cr
{0} & {e^{-ikl}}}\,\,\, ,
\end{equation}
and $T^{i}$ is the
transfer matrix of the $i^{th}$ stub.
Since all stubs are identical we have
\begin{equation}
T^{(1)}=T^{(2)}=..............=T^{(N)}=T_{a}\,\,\,   .
\end{equation}
Defining  $T_{a}M=T_{1}$, we get

\begin{equation}
T = (T_{1})^{N-1}T_{a}
\end{equation}
$T_{1}$ is not symmetric, but by a similarity transformation it
can be brought to a symmetric form. This symmetric matrix can be
diagonalized by an orthogonal matrix. Then the multiplication of
N matrices become very easy and we obtain the transfer matrix of the
entire chain which again has the form
\begin{equation}
T=\pmatrix{{1/{t_{N}}^*} & {r_{N}/t_{N}} \cr
{{r_{N}}^*/{t_{N}}^*} & {1/t_{N}}}\,\,\,  .
\end{equation}
 From this we can find $r_{N}$, the total reflection amplitude
due to N stubs
and $t_{N}$, the total transmission amplitude due to N stubs.
We have found the complex transmission amplitude $t_{N+1}$
across a periodic
structure containing (N+1) scatterers to be

\begin{equation}
1/t_{N+1}={(1-\tau)e^{ikl}\over
\tau\, 2^{N} \sqrt{\mu}}[\gamma^{N}-\delta^{N}]
+{1\over t\, 2^{N+1} \sqrt{\mu}}[\rho\, \delta^{N}-\eta\,
\gamma^{N}]\,\,\,  ,
\end{equation}
where
$$\mu={e^{-2ikl}\over t^2} + {e^{2ikl}\over t^{*2}} +2/\tau
-4\,\,\,  ,$$

$$\eta={e^{ikl}\over t^{*}}-{e^{-ikl}\over t}-\sqrt{\mu}\,\,\,,$$

$$\rho={e^{ikl}\over t^{*}}-{e^{-ikl}\over t}+\sqrt{\mu}\,\,\,,$$

$$\gamma={e^{ikl}\over t^{*}}+{e^{-ikl}\over t}+\sqrt{\mu}\,\,\,,$$

\begin{equation}
\delta={e^{ikl}\over t^{*}}+{e^{-ikl}\over t}-\sqrt{\mu}\,\,\,.
\end{equation}

Here t is the complex transmission amplitude of a single
scatterer (stub or ring) and $\tau$ is corresponding
transmission coefficient $tt^{*}$.
The total transmission coefficient $T_{N+1}$ is given by
$T_{N+1}=t_{N+1}{t_{N+1}}^{*}$.
 From the expression for the total complex
transmission amplitude one can obtain
the transmission coefficient as well as the density of states,
as discussed later. The total transmission coefficient for N+1
scatterers takes a simple form and is given by[21]

\begin{equation}
1/T_{N+1}=1+{sin^2 (N\theta)\over
sin^2(\theta)}({1\over\tau}-1)\,\,\, ,
\end{equation}
where, cos$(\theta)$=Re($e^{-ikl}$/t). Here $\theta$ is the Bloch phase
associated with the infinite periodic system.
There exists a simple relation
between the elastic scattering properties at the Fermi energy
and the conduction properties of the sample.The two terminal
conductance of an entire network g, measured in units of
$e^{2}/\pi \hbar$ and including the spin is given by
g=T, where T is the total transmission coefficient[22].

\par {\bf SERIAL ARRANGEMENT OF STUBS}
\par In the case of a single stub we can readily observe from eqn. (1)
oscillations between
zero and one in the transmission probability as a function of kL.
The wave propagating
along the stub is reflected back from the end point of the stub
(the wave function vanishes at the end point of the stub)
and gives rise to a standing wave inside the stub. This in turn
interferes with the propagating wave in the main wire. When the
phase shift kL
is $n\pi$, we
get $\tau=0$, (total backscattering)
and for phase shift of $(n+1/2)\pi$, we get
$\tau=1$ (total forward scattering).
Here $\tau$ is the transmission
probability across a single stub.
As we increase N (the
total number of stubs) to say 5, the stub lengths being the same, the
sharp drops
become extended to forbidden bands. Now we
move closer to a periodic structure like the Kronig-Penny model.
The band formation can clearly be seen in fig. (5) where we have
plotted the total transmission probability as a function of kL
for five stubs in series and we have taken L/l=1.0.
All the energy bands are identical with regions of large
transmission separated by distinct valleys.
This is a very interesting
feature characteristic of geometric scattering.
If one considers a periodic structure of
square wells or delta function potentials and study
how the band evolves as a function of the number of scatterers,
we then find striking dissimilarity[23].
There for some value of number of scatterers N say $N_{1}$
(depending upon
the strength of each potential), the first band gap becomes
well defined, the other band gaps at higher $k$ (or energy)
values are very
poorly developed or not developed at all. If we keep on
increasing the value of $N$ then for some value of
$N=N_{2}>N_{1}$ the second band gap develops but the rest are
not. A particular valley gets flatter and deeper as $N$ increases
and forms a band gap at some value of number of scatterers. To
get the complete
band for all values of k we need actually infinite number of
potential scatterers. In contrast in the
periodic
stub structures characterized by geometric scattering finite
number of stubs typically of the order of 5 stubs
are capable
of forming the entire band structure upto infinite energy of
incident electron, with wide gaps. This difference arises
mainly due to existence of transmission zeros in geometric
structures. The transmission zeros do not arise in potential
scattering problem with finite scatterers. In the case of
geometric structures the transmission coefficient oscillates
between zero and one even for arbitrarily large incident
energies, whereas in the potential scattering problem the
transmission coefficient approaches asymptotically a value of
unity as the incident energy is increased.
\par In a recent study on locally periodic potential,
saturation effect has been discussed[24]. This means the
transmission probability does not change significantly as the
number of barriers is increased. However, their observation is
based on the exploration on a limited energy range and on the
use of a gaussian wave packet that washes out the
details of the structure.
But for the case of periodic square wells and delta
function potentials it has been observed that the onset of
saturation depends on the
energy of the incident particle. For higher energies we have to
go to larger values of N to achieve the saturation. In our
mesoscopic system we find that the entire band structure for all
values of $k$ is well saturated at small value of N. There is
not much
change as we increase the value of N except that the band edges become
sharper and allowed transmission bands accommodate more
resonances. This is shown in fig. (6) where again L/$l$=1.0 and N=10.
But as we increase N the
transmission probability undergo oscillations from one (unity) to well
below one over a narrow region in the allowed bands. Each
band for N scatterers
contain N ripples or resonances if N is odd, and (N-1) ripples if
N is even. But this is true only for
the case when L is equal to l.
Out of these  resonances, one
resonance (or unit transmission) corresponds to the total
forward scattering state. This resonance always lies at the
energy corresponding to the unit transmission of a single
stub. All other resonances are of course
elementary consequences of quantum mechanical interference due
to coherent multiple scattering, and are symmetrically placed
around the value of unit transmission corresponding to a
single stub. Due to this symmetry we always find odd number of
resonances in a given band, irrespective of the total number of
scatterers being even or odd. As we increase the number of
scatterers (N), the number of spikes (or resonances) increase
proportionally within a given band and resonances come close to
each other.
Width of
the each resonance scales as 1/N and hence resonances
become sharper as their number increases with N
within a given band.
This can be
seen for $N=100$ as shown in fig. (7), where again the stub lengths
and their separations are kept the same.
\par In the above discussion we have restricted to a case where
l=L. If we fix L and increase l such that l$>$L, one
can produce more transmission resonances
in the allowed bands and make the band edges sharper. The reason is
that whenever this intermediate region (or the length l)
is such that an integral
number of
wavelengths of the incident electron fits in exactly, we get a
resonance. So, doubling the length l (i.e.,
l/L=2),within a given band,
the number of
resonances apart from the unit transmission resonance associated
with single stub, double. This increase in the
transmitted intensity at
certain places in the bands is compensated by decrease in
the transmitted intensity at certain places in the gap making
the
band edges sharper. This effect essentially arises due to the
two competing periodicities with period kl and kL (along with
their sum, difference and
harmonics). This can be seen from fig. (8), where we have plotted
the transmission coefficient versus kL for L/l=2.0 and for N=5
stubs. Comparing this fig.(8) with fig. (5) (the case in which N=5
and L/l=1.0), we notice the doubling of resonances and sharper
band edge features. In a periodic potential scatterers,
say with N delta functions, we
always get N-1 resonances in an allowed band and in contrast
their number is independent of separation between the scatterers.
\vskip.05in

In our one channel periodic case the dimensionless conductance g
takes a value between one and zero(in the forbidden band) as a function
of Fermi energy. The recent work of Leng and Lent[25] should be treated
as a generalisation of work to higher dimension(multichannel case).
For a multichannel problem the dimensionless two terminal
conductance can take a value much higher than unity depending on the
fermi energy(or number of occupied subbands). It is well known that
if the ballastic channel in higher dimension is patterned with,
constrictions or other obstuctions the quantisation of the conductance is
lost and complicated structure for conductance  emerges due to
quantum intereference and back scattering in the
channels(non adiabatic effects). However, if the ballastic
channel has a periodically modulated stucture(periodic stubs),
the quantisation in conductance is recovered but is no longer a
monotonic function of
energy. The conductance rather steps up and down between the
quantised levels some times going to zero. These new features are
sloely related to the periodic multichannel case[25].

{\bf SINGLE DEFECT IN SERIAL ARRANGEMENT OF STUBS}
\par Here we shall show how a single defect at the center of the
system (in fig. (2)) can totally alter the band structure which
keeps on changing abruptly with the strength (length) of the defect.
The length of all the stubs are assumed identical except the
central one. Again we find the total transfer matrix by
multiplying the individual transfer matrices.
If the defect stub is twice in length than the
other stubs then transmission zeros are produced exactly at the
middle of
the allowed host bands. It is easy to verify that if a single T
junction has  a transmission zero at a particular incident
energy E, then incorporating this stub in a serial one
dimensional geometric structure (ordered or disordered) produces
a transmission zero at the same energy irrespective of the
position of the stub.
This is shown in fig. (9) where we take L/$l$=1.0 and
$L_{d}/l$=2.0 where $L_{d}$ is the length of the defect stub. We
have taken
four stubs on either side of the defect stub.
We clearly observe that this single defect produces transmission
zeros in the middle of the allowed host band and also produces
resonances at the band edges.
If we make length of
the defect stub ten times larger than the length of
other stubs then
we see five zeros develop within the host
band. This is shown in fig. (10)
for L/$l$=1.0 and $L_{d}/l$=10.0 with fifty stubs on either side of
the defect. Along with the transmission zeros within a given
band we also get additional bound states in the forbidden energy
band. However, in the transmission analysis of a larger system
these bound states cannot be identified. The physics of these
bound states will be discussed later in this section.
The position of the transmission zeros within a band do not depend
in any way on whether the defect is at the center or away from the
center. Ofcourse the total transmission in general
does depend on the exact
position of the defect stub. This result is related to the
mathematical fact that transmission matrices do not commute.
The total transmission probability is related to the
interference pattern arising due to sum of infinite number of
Feynman paths which start and end at the two end points of the
system[26]. Now by having a single impurity (or impurities) in an
otherwise periodic
serial structure, all the phases of infinite Feynman
paths are altered, as
every path has to cross the impurity site at least once.
The total sum containing infinite complex amplitudes is hence
altered
non-trivially and therefore,
the total transmission probability is very
sensitive to the defect or impurity location. This fact also
implies that when the quantum nature of scatterers
become important the classical Ohm's law does not hold good
[27,28],
i.e., the total resistance defined via the transmission
coefficient is not the simple sum of individual
resistors (or scatterers). Moreover the resistance being non
additive is also a non self averaging quantity in that the
ensemble averaged
fluctuations (over all the realizations of scatterers),
in resistance dominates over the mean value of
resistance[27,28]. This is directly related to the fact that the
resistance (or the transmission coefficient) of a sample is very
sensitive to the spatial realization of scatterers. We have
shown above that one can produce a transmission zero in a
conduction band at any place with a proper choice of length of
single defect stub.
Having a zero transmission inside an allowed
band can play a very special role in band engeneering
technology. For a fixed Fermi energy by varying the length of
the single
defect stub (which can be varied by applying gate voltage)
one can thus induce a
metal insulator transition (or switch action).
\par In recent years some interest has been generated on the
possibility of positive energy
bound states above the potential barriers (in the
continuum) and their effect on the transport properties in
quantum wires[29,30]. Classically in the energy regime above the
potential barriers the
particle motion is unbounded, however, due to
quantum mechanical interference bound states can exist in this
regime. There are three different physical situations wherein
the bound states can arise. The first possibility was proposed
by von Neumann and Wigner[19]. In particular they showed that certain
spatially oscillating attractive potentials could support bound
states above the potential barriers by means of diffractive
interference. These bound states coexist with continuum states
and hence they are not robust against perturbation. A small
perturbation in the potential mixes the bound states with the
continuum states. Second class of bound states in classically
unbound region can be created in different geometries of quantum
wires having a finite cross section[29]. Several geometrical
structures like L shaped bent structures, crossed structure and
cavity structures have been considered. For example in L shaped
bent structures bound states exist, and are localized in the
bend. This is due to the fact that near the bend due to
availability of larger space the local zero point energy of the
electron is reduced considerably as compared to the zero point
energy in the side arms away from the bend. Hence the electrons
can
occupy a state (bound) at the bend which has an energy lower
than the zero point energy of the side arms, thus not able to
propagate in the arms. In contrast to the von-Neumann and Wigner
state this state does not coexist with the scattering states
(sometimes coexistence is possible due to symmetry related reasons).
Classically the particle motion is unbounded as the potential
everywhere is identically zero.
Third possibility arises whereby one can create a bound state in
a forbidden energy gap of a periodic host crystal with the help
of a single impurity potential[30].
The impurity potential is chosen
such that its resonance level lies in the energy gap of the host
crystal. Here too bound state is isolated and does not coexist
with scattering states. Such bound states have been recently
observed in quantum well structures[30]. In the following we
explicitly show that the bound state in a continuum can arise
in a one dimensional serial structure with a single defect stub.
The length of a single defect stub is so chosen that it
has a resonant state (unit transmission
state)
in the forbidden energy gap of the host serial structure.
With a few stubs
on either side of the defect stub one can spatially localize a state.
We have taken the defect stub such that $L_{d}/l$=0.75 with
only two stubs
on either sides of it and L/l=1. For this system we have
plotted the transmission coefficient versus
kL as shown in fig. (11).
In this case of a short chain there
is a peak in the transmission in the band gap. However this peak
decreases in height as we increase the length of the chain. This
is clear from the fig. (12), where we have taken 3 stubs on either
sides, with all other parameters remaining the same.
If the defect stub is such that it produces a peak in
the transmission in the band gap then a localized state will be
produced. This is because the particular mode that is allowed in the
defect stub is not allowed in the region to the right and left of it.
Thus this mode can not propagate to $\pm \infty$ and hence gets
trapped.
We can get non-vanishing transmission in
the band gap provided the localized state has some
spatial extent and does not decay appreciably on either sides of
the chain. Then it is capable of contributing to the
transmission. The peak becomes shorter if the chain length is
increased. This is because the localized state wavefunction
amplitude decays exponentially on either sides of the defect
scatterer and become
very small at the end points,
making it unable to couple wavefunctions on either
sides of the scatterer ( or localized states do not effectively
communicate with the end points of the system).
For four scatterers on either
sides, the peak completely disappears as shown in fig. (13).
It is clear that one misses the bound states in
transmission analysis
of a larger system. However, if one studies the
density of states in these systems one can easily locate the
bound states which show up as sharp peaks in the gap region.
The change in the density of states due
to scatterer, at a
particular energy is given according to Friedel theorem
as $d\theta\over(\pi dE)$, where $\theta$ is
the argument of the complex transmission amplitude[31,32].
Using this
formula and with the help of eqn. (11) we have
calculated the density of states for the system with four
scatterers on either side of the defect stub. Although
for this particular system we
do not see any peak in the transmission (see fig. (13)) at the localized
state energy, however,
we see a large peak in the density of states (see fig
(14)) at that isolated
energy showing the existence of the localized state.
Thus we have shown that one can get a positive energy
bound state in a continuum
for a serial structure by appropriately choosing the length of a
defect stub. As the potential everywhere in the network is
identically zero the classical motion of a particle is however,
unbounded for all energies.
\vfill
\eject
\par {\bf SERIAL ARRANGEMENT OF LOOPS}
\par Now on we will discuss the band formation in serial loop
structures both in the presence and in the
absence of Aharonov-Bohm flux
passing through the center of the loops.
We also study in details how the band
structure changes with the magnetic flux. The transmission
properties of
a single loop (as shown in fig. (3)) has been calculated by
Xia[15].
The transmission coefficient oscillates as a function of flux
$\phi$ with a period $\phi_{0}$(=hc/e) and is symmetric in the
flux $\phi$[33]. This is a solid state version of Aharonov-Bohm
effect. The transmission coefficient as a function of kL
(or energy) exhibits a transmission peak for certain
values of kL. This occurs whenever electron energies coincide
with the eigen energies of the entire system. In the case of two
equal arm lengths the transmission coefficient as a function of
kL does not exhibit
zeros in the absence of the magnetic field. However, in the presence
of the magnetic field through the center of the loop we observe
transmission zeros as a function of kL.
We now consider a serial structure comprising of many such loops
as shown in fig. (4). The two arms of the loop
are of equal length L each and the
length between the loops is taken to be $l$.
We follow the same procedure as for the serial stubs for
calculation of the transmission coefficient. We first consider
the situation where the magnetic field is absent. We have plotted
(fig. (15)) the transmission versus kL for N=5 loops.
The valleys in between
unit transmissions develop into wide looking band gaps for the
same reason as discussed before for serial stub structures, i.e.,
the structure moves closer to a
periodic structure like the Kronnig Penny model. The number
resonances in the conduction band depend on the lengths L and l.
The transmission is not zero in the gaps but saturation sets in
identically for all gaps. Bands are
clearly formed for $N=5$ as shown
in fig. (15) where L/$l$=1.0.
The bands and the band gaps become more
defined as we increase the value of N. It is interesting to
note that for a serial loop structure the first allowed
conduction band starts right from the energy zero. In contrast,
in the serial stub case we always find a forbidden gap around
energy zero. This follows from the fact that for an isolated
loop the lowest eigenstate in the absence of magnetic field has
momentum k=0.
Instead of L/$l$=1.0, if we take $l$ to be smaller such that
L/$l$=100.0(say), with N
remaining the same then the band structure shows an
effect opposite to that of a serially arranged potentials. We find
that the higher band gaps are formed more easily and saturate
faster as N increases. This is shown in fig. (16). This happens
because if the electron
wavelength is larger than separation between loops then multiple
reflection between two loops cannot form standing waves.
Only high energy electrons
with wavelength much smaller than the separation between loops are
strongly affected by these regions due to multiple scattering
and the effect of periodicity is felt.

We now consider the evolution of the band structure as a
function of the magnetic flux in the serial loop structure.
In the presence of the magnetic flux
through each loop each band splits into two (except the lowest
one) and
band gap appears at the center of each band. This is shown in fig. (17),
where we have taken $\alpha=.2$ and L/$l$=1.0 for N=5 loops.
This band gap that develops due to the magnetic field can also be
made to develop at other places of the band than the center by
choosing a different value of L/$l$. This is seen in fig. (18) where
L/$l$=100.0, $\alpha$=0.6 and N=20.
The band structure evolves continuously as we change the enclosed
magnetic flux and the band gap increases continuously with the
magnetic flux. In fig. (19) the dashed curve is for N=5.0,
L/l=1.0 and $\alpha$=.4 whereas the solid curve is for N=5.0,
L/l=1.0 and
$\alpha$=0.2. The band gap is seen to be more for the larger
magnetic flux.
In this way these special band gaps initially keep widening with
$\alpha$ upto $\alpha=\pi$ where the transmission is identically
zero for all values of kL as
evidient from eqn. (2), and then it starts narrowing with
$\alpha$, finally disappearing at $\alpha=2\pi$.
If we increase
$\alpha$ further then the same cycle is repeated, with a period
$\phi_{0}$. These band gaps appear with the magnetic field
because when flux through a loop
is non zero then the transmission probability across a single loop
show zeros. Each maxima of unit transmission splits into two and a
zero develops in between.
This is because the magnetic field breaks the time
reversal symmetry. For
an isolated closed loop (ring)
the eigenstates in the absence of magnetic field
can be calculated via
the boundary condition
$e^{i2kL}=1,$
where $k$ is the wave vector and $2L$ is the
circumference of the loop. So
$k_{n}={n\pi\over 2L}, n=0,\pm 1,\pm 2....\,\,\,\,\,,$
and the discrete energy states are given by
$E_{n}={\hbar^{2}{k_{n}}^{2}\over 2m}$.
The states with $n=\pm 1$ are degenerate.Similarly $n=\pm 2$
states are degenerate and so on. Now when we connect perfect
leads on two
sides, the transmission probability exhibits a peak transmission
(T=1) for certain values of kL. This happens whenever the
incident electron energy coincides with one of the eigen
energies of the system. The deviations from the values of the
exact energy states of the closed ring follow from the
additional junction scattering due to the leads.
Each such peak is degenerate.
But once the flux through the isolated closed loop is non-zero this
degeneracy
is removed and resonance peak splits into two peaks
and a zero in the transmission develops in between. The separation
between the peaks is periodic in flux reaching the maximum value at
$\alpha=\pi$ and the transmission across a single loop is
identically zero for all values of kL.
In periodic structure the sharp zeros develop
into wide looking band gaps and hence one
can develop band gaps at the middle of the
bands with the help of magnetic field. It is important to know
that as we tune the magnetic field the bands shift their
position on the energy axis.
The
magnetic field destroys the time reversal symmetry and as a
consequence degeneracy of the states carrying current clockwise
and anticlockwise in the loops is lifted. Depending on the Fermi
energy uncompensated current flows in either of the directions.
In the case of the isolated ring the persistent current
oscillates from one state to the next and the total persistent
current is given by a sum due to all occupied levels[34-36]. The
current $i_{n}$ in the loop carried by $n^{th}$ eigenstate with
energy $E_{n}(\phi)$ is proportional to
$(-1/c){\partial E_{n}(\phi)\over \partial \phi}$.
In the case of the periodic loop structure considered here all the
odd bands (1,3, etc.)
initially narrow as we increase the
magnetic field.
In these bands the resonances at lower energy
shift right on the energy scale
and the resonances on the higher energy side of
the band shift to the left. This in turn implies that the
band exhibits a mixed magnetic behavior,
in that the lower energy eigen states carry a diamagnetic current
whereas the higher energy eigenstates
within these bands carry paramagnetic currents.
The even bands (2,4, etc.) always shift to the left on the
energy axis initially as we increase magnetic field.
This amounts to a fact
that all energy eigenstates in an even band
carry a paramagnetic current.
In fig. (19) we have shown the evolution of second and third
bands for two values of the magnetic field.
Due to the band formation we expect to observe
larger equilibrium magnetic response in the loop structures as
compared to isolated rings. Since the total persistent current
is due to the sum over all states and the nature of persistent
current within a paramagnetic band does not change, i.e.,
within a given paramagnetic band all the energy levels carry
paramagnetic current,
by appropriately varying the Fermi level in the band one
can obtain larger contribution to the persistent current.
As in the case for a serial stub structures, having a single loop
with different dimensions (circumference) from the rest, one can
tailor the band structure and one can also create localized states
in the continuum. The physics is essentially the same.

{\bf CONCLUSIONS}

We have analysed the problem of band formation in a periodic
multiple stub and loop(in the presence of magnetic flux)
configurations, using quantum wave guide theory on networks.
Throughout we have restricted ourselves to a single channel
free electron network. In these network the scattering is
solely due to the geometric nature of the problem. The evolution of
band formation  as a function of number of scatterers is
compared with a conventional one dimensional potential
scattering system and several notable differences have been
pointed out. A single defect in  an otherwise periodic system
modifies band properties nontrivially and argued that this
fact is a consequence of the existence of zeroes in geometric
structures. The sensitivity of the band strucutre to a single
defect can be utilised for band tailoring and
for quantum device operations. We have also shown that by a
proper choice of a single defect one can produce bound states in
continuum in the sense of vonNuemann and Wigner. Finally we have also
discussed the magnetic properties arising in loop structures
in the presence of Aharonov Bohm fulx. In paricular we explained
notion of diamagnetic and mixed properties of  bands due to orbital
currents.

\end{document}